\documentclass[11pt]{JHEP3}
\usepackage{amsmath}
\usepackage{amssymb}
\usepackage[numbers,sort&compress]{natbib}
\newcommand{\Sinf}{S_{\infty}}
\newcommand{\ABA}{\mbox{\tiny ABA}}
\newcommand{\Wrap}{\mathrm{wrap}}

\newcommand{\Op}{\mathcal{O}}

\newcommand{\cN}{\mathcal{N}}

\def\HBS{\mathbb S}
\def\HS{S}
\def\reciP{\mathcal{P}}

\def\cP{\mathcal{P}}
\def\cT{\mathcal{T}}

\newcommand{\beq}{\begin{equation}}
\newcommand{\eeq}{\end{equation}}
\newcommand{\beqa}{\begin{eqnarray}}
\newcommand{\eeqa}{\end{eqnarray}}
\newcommand{\M}{M}

\def\ztw{\zeta_2}
\def\zt{\zeta_3}
\def\zfr{\zeta_4}
\def\zf{\zeta_5}
\def\zsx{\zeta_6}
\def\zs{\zeta_7}

\title{Twist-2 at five loops: Wrapping corrections without wrapping computations}

\author{V.~N.~Velizhanin
\\
\\
Institut f{\"u}r  Mathematik und Institut f{\"u}r Physik\\
Humboldt-Universit{\"a}t zu Berlin\\
IRIS Adlershof, Zum Gro\ss{}en Windkanal 6\\
12489 Berlin, Germany\\
\\
Theoretical Physics Division\\
Petersburg Nuclear Physics Institute\\
Orlova Roscha, Gatchina\\
188300 St.~Petersburg, Russia\\
\\
E-mail: \email{velizh@physik.hu-berlin.de}
}

\abstract{
Using known all-loop results from the BFKL and generalized double-logarithmic equations and large spin limit we have recomputed the five-loop planar anomalous dimension of twist-2 operators without consideration of any wrapping effects.
One part of the anomalous dimension was calculated in an usual way with the help of Asymptotic Bethe Ansatz.
The rest part, related with the wrapping effects, was reconstructed from a known constraints with the help of methods from
the numbers theory.
}

\preprint{
          \tiny{HU-Mathematik-2013-21}\\[-.15ex]
          \tiny{HU-EP-13/73} %\\[-.5ex]
          }
\begin{document}

\maketitle

\section{Introduction}
\label{sec:intro}

During the investigation of AdS/CFT-correspondence~\cite{Maldacena:1997re,Gubser:1998bc,Witten:1998qj} it was found that the anomalous dimension of BMN-operators~\cite{Berenstein:2002jq} in $\cN=4$ SYM theory in the leading order of perturbative theory can be calculated with the help of integrability~\cite{Minahan:2002ve}\footnote{Earlier, the similar integrability was opened in Quantum Chromodynamics in the Regge limit~\cite{Lipatov:1993yb,Lipatov:1994xy,Faddeev:1994zg} and for some of operators~\cite{Braun:1998id}.}.
A generalization to the higher orders together with the investigations of the integrable structures from the superstring theory side, started in ref.~\cite{Bena:2003wd}, allowed to formulate all-loop asymptotic Bethe ansatz (ABA) equations
\cite{Beisert:2003tq,Beisert:2003yb,Beisert:2003jj,Beisert:2003ys,Serban:2004jf,Kazakov:2004qf,Beisert:2004hm,Arutyunov:2004vx,Staudacher:2004tk,
Beisert:2005fw,Beisert:2005tm,Janik:2006dc,Hernandez:2006tk,Arutyunov:2006iu,Beisert:2006ib,Eden:2006rx,Bern:2006ew,Beisert:2006ez}.
The next step in the studying of integrability became the investigation of a wrapping effects for the operators with the finite length, for which ABA gives an incomplete result~\cite{SchaferNameki:2006ey,Kotikov:2007cy}.
Initially, for the computation of the wrapping corrections it was suggested to use a results from the string theory side, where the wrapping corrections correspond to the effects of finite volume in which a string theory is considered (a generalization~\cite{Ambjorn:2005wa} of L\"uscher formulae~\cite{Luscher:1985dn,Luscher:1986pf}).
Expanding the obtained exact result for the modification of the energy of string state under perturbative theory it is possible to calculate the wrapping corrections for the anomalous dimension of the corresponding finite size operators~\cite{Bajnok:2008bm,Bajnok:2008qj,Bajnok:2009vm,Lukowski:2009ce}.
Firstly this method was applied for the simplest operator in $\cN=4$ SYM theory, i.e. for Konishi operator in the forth order of perturbative theory, where the finite size effect appears from the first time~\cite{Bajnok:2008bm}.
Earlier the similar result was obtained from the perturbative calculations~\cite{Fiamberti:2007rj,Fiamberti:2008sh} (based on the method from ref.~\cite{Sieg:2005kd}), which coincided with the result of ref.~\cite{Bajnok:2008bm} after a minor corrections. Slightly later a full direct four-loop diagrammatic calculations from the field theory side were performed~\cite{Velizhanin:2008jd,Velizhanin:2008pc} and the obtained result coincided with both previous. Then, this method was generalized to the twist-2 operators with an arbitrary Lorentz spin $\M$ (arbitrary number of covariant derivatives $D$ in operator $ZD^{\M}Z$) in the same order~\cite{Bajnok:2008qj}. The obtained result was found in a full agreement with the all-loop predictions, coming from the Balitsky-Fadin-Kuraev-Lipatov (BFKL) equation~\cite{Lipatov:1976zz,Kuraev:1977fs,Balitsky:1978ic} and from the result of the direct perturbative calculations of the leading transcendentality contribution~\cite{Velizhanin:2008pc}, which serves as an important test of the correctness of the used methods. Then, L\"uscher corrections were expanded to the next order for the calculation of the anomalous dimension of Konishi operator~\cite{Bajnok:2009vm} at five loops and for the anomalous dimension of the twist-2 operators with the arbitrary Lorentz spin in the same order~\cite{Lukowski:2009ce}. The last result was again in the full agreement with the predictions, coming from the BFKL equation\footnote{Obtained formulae for the leading and next-to-leading wrapping corrections were also applied for the calculations of the five- and six-loop anomalous dimensions of the twist-3 operator~\cite{Beccaria:2009eq,Velizhanin:2010cm}, for which wrapping effects appear in one order more with compare to twist-2 operator.}.

In the same time the investigations of the more general integrable system were performed, which can be applied for the computations of the anomalous dimension of any operators in any order of perturbative theory along the approach, proposed by A.~Zamolodchikov~\cite{Zamolodchikov:1989cf,Zamolodchikov:1991et}. In the result of such investigations the set of TBA equations and the Y-system were formulated~\cite{Gromov:2009tv,Gromov:2009bc,Gromov:2009zb,Bombardelli:2009ns,Arutyunov:2009ax,Arutyunov:2009ur,Arutyunov:2010gb,Balog:2010xa,Balog:2010vf,Bajnok:2010ud}. Both methods were tested with reproduction of the known four- and five-loop results and allow to obtain a lot of new results not only in $\cN=4$ SYM theory, but in other integrable models. However further test of the developed methods is based mainly on a self-consistency and it would be nice to have some independent crosscheck. One of such important test can come from the analysis of the analytical properties  of the full anomalous dimension of twist-2 operators, for which a lot of information is exist. All these predictions are based on a similar results, obtained earlier from the studying of the properties of twist-2 operators in QCD, where they give the leading contribution through the application of an operator product expansion to the deep-inelastic processes, actively studied experimentally. Initially, for these purposes were used all-loop predictions, coming from the BFKL and double-logarithmic equations~\cite{Gorshkov:1966ht,Gorshkov:1966hu,Gorshkov:1966qd,Kirschner:1982qf,Kirschner:1982xw,Kirschner:1983di}. But some times ago we found, that in $\cN=4$ SYM theory there is a rather simple generalization of double-logarithmic equation~\cite{Velizhanin:2011pb}, which give a lot of constraints on the analytical structure of the anomalous dimension of twist-2 operators in any-loop orders. The presence a lot of information about analytical properties of twist-2 anomalous dimension led to the idea to use this information not for the testing of an obtained results, but directly for the computations of the general form of the anomalous dimension of twist-2 operator for the arbitrary Lorentz spin, i.e. for the reconstruction of the wrapping corrections. From the earlier obtained results in QCD and in $\cN=4$ SYM theory it is well known, that the anomalous dimension of twist-2 operators are expressed through the nested harmonic sums, defined as (see~\cite{Vermaseren:1998uu}):
\beq \label{vhs}
\HS_a (\M)=\sum^{\M}_{j=1} \frac{(\mbox{sgn}(a))^{j}}{j^{\vert a\vert}}\, , \qquad
\HS_{a_1,\ldots,a_n}(\M)=\sum^{\M}_{j=1} \frac{(\mbox{sgn}(a_1))^{j}}{j^{\vert a_1\vert}}
\,\HS_{a_2,\ldots,a_n}(j)\, .
\eeq
In other words the nested harmonic sums form the finite basis for the anomalous dimension of the twist-2 operators in each order of perturbative theory.
According to the principle of a maximal transcendentality~\cite{Kotikov:2002ab} (see also~\cite{Kotikov:2003fb,Kotikov:2004er}) at the $\ell$-th order of the perturbation theory the anomalous dimension of twist-2 operators is expressed through the nested harmonic sums of the order $(2\ell-1)$, or through the products of zeta functions and harmonic sums for which the sum of the arguments of the zeta functions and the orders of the harmonic sums is equal to $(2\ell-1)$. For the fixing the coefficients in the ansatz, obtained from the corresponding basis, it is necessary to use all information, which we have in this order of perturbative theory. For the reconstruction of the general expression for the anomalous dimension of twist-2 operators, coming from the ABA one can used results for the definite values of $\M$. For the full anomalous dimension besides analytical properties we can used a generalized Gribov-Lipatov relation~\cite{Dokshitzer:2005bf,Dokshitzer:2006nm} and looking for a reciprocity-respecting function, instead of the anomalous dimension, which is related to each other through:
\beq \label{Pfunction}
\gamma(\M) = \reciP \left(\M+\frac{1}{2} \gamma(\M) \right)\,.
\eeq
This allow considerably reduce the number of the harmonic sums in the basis, because these sums (or a combinations of sums) should respect this symmetry. We found, that more suitable move from the nested harmonic sums to a binomial harmonic sums, which are definite as (see \cite{Vermaseren:1998uu}):
\beq\label{BinSums}
\HBS_{i_1,\ldots,i_k}(N)=(-1)^N\sum_{j=1}^{N}(-1)^j\binom{N}{j}\binom{N+j}{j}\HS_{i_1,...,i_k}(j)\,,
\eeq
where $\HS_{i_1, \ldots ,i_k}$ are the nested harmonic sums (\ref{vhs}) and all indices $i_k$ are always positive. Note again, that the number of the binomial harmonic sums is much less, than the number of the usual nested harmonic sums.

In the fourth order of perturbative theory an available information is enough and the coefficients in ansatz are fixed uniquely from all know information (see \cite{Velizhanin:2011pb}). Unfortunately, the number of sums in ansatz grows with the order of perturbative theory faster than the number of constraints, coming from the know all-loop results, so, the number of an unknown coefficients in ansatz are more than the number of available equations. However, in such case we can use the method, which we already used for the reconstruction of the six-loop anomalous dimension of the twist-3 operators in $\cN=4$ SYM theory~\cite{Velizhanin:2010cm} and for the anomalous dimension of the non-singlet transverse twist-2 operators in QCD in third order of perturbative theory~\cite{Velizhanin:2012nm}. This method based on the fact, that the sought-for coefficients in ansatz should be an integer numbers and, more over, some of them should be zero. With these conditions we obtain the system of a linear Diophantine equations, which can be solved with the different method from the numbers theory. In our previous papers we used LLL-algorithm~\cite{Lenstra:1982} to find a vector, which can solve an initial system. But we have found also other method, which will be applied for the calculations of the five-loop anomalous dimension of the twist-2 operators, presented in this paper.

The paper organized in the following way. In the Section~\ref{sec:fiveloop} we give the result for the five-loop planar anomalous dimension of the twist-2 operator, coming from the ABA, which was obtained earlier in ref.~\cite{Lukowski:2009ce}. In the Section~\ref{sec:weak} we obtain all expressions, which we will use then for the reconstruction of the full planar anomalous dimension. The Section~\ref{sec:L5} is devoted to the reconstruction of the five-loop planar anomalous dimension, which was computed earlier in ref.~\cite{Lukowski:2009ce}, from the constraints obtained in the Section~\ref{sec:weak}.
% ADD, which was computed earlier in ref.~\cite{Lukowski:2009ce}.
We will apply two methods and will give a simple example, which shows how the methods are work. In the Conclusion we summarize our results and the overall methods.
% ADD
The Appendix contains a general expression for the next-to-leading order wrapping correction to the anomalous dimension of twist-2 operators in the $\beta$-deformed $\cN=4$ SYM theory from the results of ref.~\cite{deLeeuw:2010ed} reconstructed with the help of LLL-algorithm.

\section{Five-loop anomalous dimension from Bethe Ansatz}
\label{sec:fiveloop}

We start with the part of the anomalous dimension, which can be calculated with the help of asymptotic Bethe ansatz~\cite{Beisert:2005fw,Beisert:2006ez}. The result at five-loop order  together with the detailed description of the computations can be found in ref.~\cite{Lukowski:2009ce}. Here we present only the final result in the terms of the reciprocity-respecting functions~(\ref{Pfunction}) and the binomial harmonic sums~(\ref{BinSums}).
Substituting the perturbative expansion for the anomalous dimension
\begin{equation}
\gamma^{\ABA}(M)=\sum_{l=1}^\infty g^{2l}\,\gamma^{\ABA}_{2l}(M)\,.\label{gABA}
\end{equation}
in eq.~(\ref{Pfunction}), one finds
\begin{equation}
\reciP^{\ABA}(M)=\sum_{l=1}^\infty g^{2l}\,\reciP^{\ABA}_{2l}(M)\,.\label{PABA}
\end{equation}
From eqs.~(\ref{Pfunction}) and~(\ref{PABA}) one can find, that at five-loop order the reciprocity function $\reciP_{10}$ is related with the anomalous dimension $\gamma_{10}$ (see Appendix B of ref.~\cite{Beccaria:2009eq}):
\beqa
\reciP_{10}(M)&=&\reciP_{10}\ =\
\reciP_{10}^{\rm rational}
+\reciP_{10}^{\zt}\zt
+\reciP_{10}^{\zf}\zf
%+\reciP_{10}^{\zs}\zs
\,,\\[3mm]
\reciP_{10}^{\rm rational}&=&
\gamma_{10}^{\rm rational}
-\frac{1}{4} \left(
%\gamma_6^2
\, \gamma_4 \gamma_6
+\, \gamma_2 \gamma_{8}^{\rm rational}
\right)'
%\nonumber\\&&
+\frac{1}{32} \left(
%\gamma_4^3
\, \gamma_2 \gamma_4^2
+\, \gamma_2^2 \gamma_6
\right)''
\nonumber\\&&
-\frac{1}{384} \left(\, \gamma_2^3 \gamma_4\right)'''
+\frac{1}{30720}\left(\gamma_2^5\right)''''\,,
\label{P10}\\[3mm]
\reciP_{10}^{\zt}&=&
\gamma_{10}^{\zt}
-\frac{1}{4} \left(\gamma_2 \gamma_8^{\zt}\right)'\,,
\label{Pzt}\\[3mm]
\reciP_{10}^{\zf}&=&
\gamma_{10}^{\zf}\,,
\label{Pzf}
\eeqa
where each prime marks derivative over $\M$.
The result for $\reciP_{10}^{\rm rational}(\M)$ is equal to~\cite{Lukowski:2009ce}:
\begin{table}[ht]
\beqa
\frac{\reciP^{\textrm{rational}}_{10}}{128}&=&
-5 \, \HBS_{2,2,5}
-\, \HBS_{2,6,1}
+19 \, \HBS_{3,1,5}
-20 \, \HBS_{3,2,4}
+21 \, \HBS_{4,1,4}
-24 \, \HBS_{4,2,3}
+25 \, \HBS_{5,1,3}\nonumber\\ & &
-18 \, \HBS_{5,2,2}
+7 \, \HBS_{6,1,2}
-4 \, \HBS_{6,2,1}
-2 \, \HBS_{1,1,2,5}
+2 \, \HBS_{1,1,6,1}
-2 \, \HBS_{1,2,1,5}
-\, \HBS_{1,2,2,4}\nonumber\\ & &
+\, \HBS_{1,2,3,3}
+\, \HBS_{1,2,4,2}
-6 \, \HBS_{1,2,5,1}
+23 \, \HBS_{1,3,1,4}
-24 \, \HBS_{1,3,2,3}
-\, \HBS_{1,3,4,1}
+23 \, \HBS_{1,4,1,3}\nonumber\\ & &
-20 \, \HBS_{1,4,2,2}
-\, \HBS_{1,4,3,1}
+13 \, \HBS_{1,5,1,2}
-12 \, \HBS_{1,5,2,1}
+6 \, \HBS_{1,6,1,1}
-2 \, \HBS_{2,1,1,5}
+5 \, \HBS_{2,1,2,4}\nonumber\\ & &
+\, \HBS_{2,1,3,3}
+\, \HBS_{2,1,4,2}
-5 \, \HBS_{2,1,5,1}
-16 \, \HBS_{2,2,1,4}
+17 \, \HBS_{2,2,2,3}
-2 \, \HBS_{2,2,3,2}
+14 \, \HBS_{2,2,4,1}\nonumber\\ & &
-29 \, \HBS_{2,3,1,3}
+25 \, \HBS_{2,3,2,2}
+4 \, \HBS_{2,3,3,1}
-19 \, \HBS_{2,4,1,2}
+20 \, \HBS_{2,4,2,1}
-12 \, \HBS_{2,5,1,1}
+20 \, \HBS_{3,1,1,4}\nonumber\\ & &
-22 \, \HBS_{3,1,2,3}
-8 \, \HBS_{3,1,3,2}
+6 \, \HBS_{3,1,4,1}
-26 \, \HBS_{3,2,1,3}
+36 \, \HBS_{3,2,2,2}
-5 \, \HBS_{3,2,3,1}
-6 \, \HBS_{3,3,1,2}\nonumber\\ & &
+5 \, \HBS_{3,3,2,1}
-2 \, \HBS_{3,4,1,1}
+22 \, \HBS_{4,1,1,3}
-24 \, \HBS_{4,1,2,2}
+6 \, \HBS_{4,1,3,1}
-18 \, \HBS_{4,2,1,2}
+18 \, \HBS_{4,2,2,1}\nonumber\\ & &
-2 \, \HBS_{4,3,1,1}
+14 \, \HBS_{5,1,1,2}
-10 \, \HBS_{5,1,2,1}
-14 \, \HBS_{5,2,1,1}
+8 \, \HBS_{6,1,1,1}
+4 \, \HBS_{1,1,1,1,5}
-6 \, \HBS_{1,1,1,2,4}\nonumber\\ & &
-2 \, \HBS_{1,1,1,3,3}
-2 \, \HBS_{1,1,1,4,2}
+6 \, \HBS_{1,1,1,5,1}
-4 \, \HBS_{1,1,2,1,4}
+4 \, \HBS_{1,1,2,2,3}
+5 \, \HBS_{1,1,2,3,2}\nonumber\\ & &
-13 \, \HBS_{1,1,2,4,1}
+24 \, \HBS_{1,1,3,1,3}
-20 \, \HBS_{1,1,3,2,2}
-5 \, \HBS_{1,1,3,3,1}
+16 \, \HBS_{1,1,4,1,2}
-19 \, \HBS_{1,1,4,2,1}\nonumber\\ & &
+12 \, \HBS_{1,1,5,1,1}
-4 \, \HBS_{1,2,1,1,4}
+7 \, \HBS_{1,2,1,2,3}
+4 \, \HBS_{1,2,1,3,2}
-8 \, \HBS_{1,2,1,4,1}
-19 \, \HBS_{1,2,2,1,3}\nonumber\\ & &
+9 \, \HBS_{1,2,2,2,2}
+24 \, \HBS_{1,2,2,3,1}
-22 \, \HBS_{1,2,3,1,2}
+31 \, \HBS_{1,2,3,2,1}
-22 \, \HBS_{1,2,4,1,1}
+22 \, \HBS_{1,3,1,1,3}\nonumber\\ & &
-24 \, \HBS_{1,3,1,2,2}
+6 \, \HBS_{1,3,1,3,1}
-20 \, \HBS_{1,3,2,1,2}
+23 \, \HBS_{1,3,2,2,1}
-6 \, \HBS_{1,3,3,1,1}
+16 \, \HBS_{1,4,1,1,2}\nonumber\\ & &
-11 \, \HBS_{1,4,1,2,1}
-22 \, \HBS_{1,4,2,1,1}
+16 \, \HBS_{1,5,1,1,1}
-4 \, \HBS_{2,1,1,1,4}
+7 \, \HBS_{2,1,1,2,3}
+4 \, \HBS_{2,1,1,3,2}\nonumber\\ & &
-8 \, \HBS_{2,1,1,4,1}
+3 \, \HBS_{2,1,2,1,3}
-11 \, \HBS_{2,1,2,2,2}
+14 \, \HBS_{2,1,2,3,1}
-9 \, \HBS_{2,1,3,1,2}
+16 \, \HBS_{2,1,3,2,1}\nonumber\\ & &
-12 \, \HBS_{2,1,4,1,1}
-16 \, \HBS_{2,2,1,1,3}
+13 \, \HBS_{2,2,1,2,2}
+5 \, \HBS_{2,2,1,3,1}
+28 \, \HBS_{2,2,2,1,2}
-67 \, \HBS_{2,2,2,2,1}\nonumber\\ & &
+32 \, \HBS_{2,2,3,1,1}
-22 \, \HBS_{2,3,1,1,2}
+15 \, \HBS_{2,3,1,2,1}
+31 \, \HBS_{2,3,2,1,1}
-23 \, \HBS_{2,4,1,1,1}
+20 \, \HBS_{3,1,1,1,3}\nonumber\\ & &
-22 \, \HBS_{3,1,1,2,2}
+6 \, \HBS_{3,1,1,3,1}
-8 \, \HBS_{3,1,2,1,2}
+2 \, \HBS_{3,1,2,2,1}
+6 \, \HBS_{3,1,3,1,1}
-20 \, \HBS_{3,2,1,1,2}\nonumber\\ & &
+15 \, \HBS_{3,2,1,2,1}
+25 \, \HBS_{3,2,2,1,1}
-7 \, \HBS_{3,3,1,1,1}
+16 \, \HBS_{4,1,1,1,2}
-11 \, \HBS_{4,1,1,2,1}
-11 \, \HBS_{4,1,2,1,1}\nonumber\\ & &
-23 \, \HBS_{4,2,1,1,1}
+16 \, \HBS_{5,1,1,1,1}
-4 \, \HBS_{1,1,1,1,2,3}
-4 \, \HBS_{1,1,1,1,3,2}
+8 \, \HBS_{1,1,1,1,4,1}
+4 \, \HBS_{1,1,1,2,2,2}\nonumber\\ & &
-13 \, \HBS_{1,1,1,2,3,1}
+18 \, \HBS_{1,1,1,3,1,2}
-25 \, \HBS_{1,1,1,3,2,1}
+16 \, \HBS_{1,1,1,4,1,1}
+3 \, \HBS_{1,1,2,1,2,2}\nonumber\\ & &
-5 \, \HBS_{1,1,2,1,3,1}
-20 \, \HBS_{1,1,2,2,1,2}
+53 \, \HBS_{1,1,2,2,2,1}
-33 \, \HBS_{1,1,2,3,1,1}
+18 \, \HBS_{1,1,3,1,1,2}\nonumber\\ & &
-11 \, \HBS_{1,1,3,1,2,1}
-25 \, \HBS_{1,1,3,2,1,1}
+20 \, \HBS_{1,1,4,1,1,1}
+3 \, \HBS_{1,2,1,1,2,2}
-5 \, \HBS_{1,2,1,1,3,1}\nonumber\\ & &
+14 \, \HBS_{1,2,1,2,2,1}
-17 \, \HBS_{1,2,1,3,1,1}
-18 \, \HBS_{1,2,2,1,1,2}
+11 \, \HBS_{1,2,2,1,2,1}
+52 \, \HBS_{1,2,2,2,1,1}\nonumber\\ & &
-27 \, \HBS_{1,2,3,1,1,1}
+16 \, \HBS_{1,3,1,1,1,2}
-11 \, \HBS_{1,3,1,1,2,1}
-11 \, \HBS_{1,3,1,2,1,1}
-27 \, \HBS_{1,3,2,1,1,1}\nonumber\\ & &
+20 \, \HBS_{1,4,1,1,1,1}
+3 \, \HBS_{2,1,1,1,2,2}
-5 \, \HBS_{2,1,1,1,3,1}
+14 \, \HBS_{2,1,1,2,2,1}
-17 \, \HBS_{2,1,1,3,1,1}\nonumber\\ & &
+16 \, \HBS_{2,1,2,2,1,1}
-11 \, \HBS_{2,1,3,1,1,1}
-16 \, \HBS_{2,2,1,1,1,2}
+11 \, \HBS_{2,2,1,1,2,1}
+11 \, \HBS_{2,2,1,2,1,1}\nonumber\\ & &
+45 \, \HBS_{2,2,2,1,1,1}
-27 \, \HBS_{2,3,1,1,1,1}
+16 \, \HBS_{3,1,1,1,1,2}
-11 \, \HBS_{3,1,1,1,2,1}
-11 \, \HBS_{3,1,1,2,1,1}\nonumber\\ & &
-11 \, \HBS_{3,1,2,1,1,1}
-27 \, \HBS_{3,2,1,1,1,1}
+20 \, \HBS_{4,1,1,1,1,1}
-16 \, \HBS_{1,1,1,1,2,2,1}
+24 \, \HBS_{1,1,1,1,3,1,1}\nonumber\\ & &
-28 \, \HBS_{1,1,1,2,2,1,1}
+20 \, \HBS_{1,1,1,3,1,1,1}
-20 \, \HBS_{1,1,2,2,1,1,1}
+20 \, \HBS_{1,1,3,1,1,1,1}\nonumber\\ & &
-20 \, \HBS_{1,2,2,1,1,1,1}
+20 \, \HBS_{1,3,1,1,1,1,1}
-20 \, \HBS_{2,2,1,1,1,1,1}
+20 \, \HBS_{3,1,1,1,1,1,1}\,.
\eeqa
\caption{The five-loop function $\reciP_{10}^{\rm rational} (\M)$.} \label{pfiveloop}
\end{table}
\clearpage
\newpage
\noindent The functions $\reciP^{\zt}_{10}$ and $\reciP^{\zf}_{10}$ look like~\cite{Beccaria:2009rw,Lukowski:2009ce}
\beqa
\frac{\reciP_{10}^{\zt}}{256}&=&
3 \HBS_{1,5}
-4 \HBS_{2,4}
-\HBS_{3,3}
-\HBS_{4,2}
+3 \HBS_{5,1}
+2 \HBS_{1,1,4}
-4 \HBS_{1,2,3}
-2 \HBS_{1,3,2}
+2 \HBS_{1,4,1}\nonumber\\
&&-\HBS_{2,1,3}
+5 \HBS_{2,2,2}
-2 \HBS_{2,3,1}
-2 \HBS_{3,2,1}
+2 \HBS_{4,1,1}
-3 \HBS_{1,1,2,2}
+\HBS_{1,1,3,1}
-\HBS_{1,2,1,2}\nonumber\\
&&+2 \HBS_{1,3,1,1}
-\HBS_{2,1,1,2}
+\HBS_{2,1,2,1}
+\HBS_{3,1,1,1}\,,\label{ABAz3}\\
%\eeqa
%\beqa
\frac{\reciP_{10}^{\zf}}{640}&=&\HBS_1 (\HBS_{2,1}-\HBS_3)\,.\label{ABAz5}
\eeqa
\section{Weak-coupling constraints}\label{sec:weak}

In this section we will write down the known weak-coupling constraints on the five-loop anomalous dimension of twist-2 operators in $\cN=4$ SYM theory. We will use three classes of constraints, which are provided by the BFKL equation, by the generalized double-logarithmic equation and by the large $\M$ limit.

\subsection{BFKL equation} \label{sec:BFKL}

The relation between the anomalous dimension of twist-2 operators and the Balitsky-Fadin-Kuraev-Lipatov (BFKL) equation \cite{Lipatov:1976zz,Kuraev:1977fs,Balitsky:1978ic} and its next-to-leading logarithm approximation (NLLA) generalization \cite{Fadin:1998py,Kotikov:2000pm}
 emerges upon an analytic continuation of the function $\gamma(g,M)$ to complex values of $\M$. This is straightforward in the one-loop case since
\beq
\gamma_{2}(M) = 8\,g^2\,S_1 (M) = 8\,g^2\, \left(\Psi(M+1)-\Psi(1)\right)\,,
\eeq
where $\Psi(x)=\frac{d}{dx}\,\log \Gamma(x)$ is the digamma function.
At any loop order one expects a singularities at all {negative integer} values of  $\M$. The first in this series of singular points,
\beq
\label{omega}
M=-1+\omega\,,
\eeq
corresponds to the BFKL pomeron. In the above formula $\omega$ should be considered infinitesimally small. The BFKL equation predicts that, if expanded in $g$, the $\ell$-loop anomalous dimension $\gamma_{2 \ell} (\omega)$ exhibits poles in $\omega$. Moreover, the residues and the order of the poles can be derived directly from the BFKL equation.
Up to the next-to-leading logarithm approximation the BFKL equation for twist-2 operators in $\cN=4$ SYM theory in the dimensional reduction scheme can be written as follows~\cite{Fadin:1998py,Kotikov:2000pm}
\beq
\frac{\omega}{-4\,g^2} = \chi (\gamma )-g^2\,\delta (\gamma )\,,
\eeq
where
\beqa
\label{gammanlo}
\chi (\gamma ) &=&
\Psi\left(-\frac{\gamma}{2}\right)+\Psi\left(1+\frac{\gamma}{2}\right)-
2\,\Psi\left(1\right)\, ,\\[4mm]
\delta (\gamma ) &=&4\,\chi ^{\,\prime \prime } (\gamma )
+6\,\zeta(3)+2\,\zeta(2)\,\chi (\gamma )+4\,\chi (\gamma )\,\chi ^{\,\prime} (\gamma )  \nonumber \\[2mm]
& & -\frac{\pi^3}{\sin \frac{\pi \gamma}{2}}- 4\,\Phi \left(-\frac{\gamma}{2}
\right) -4\,\Phi \left(1+\frac{\gamma}{2} \right)\,.
\eeqa
The function $\Phi (\gamma )$ is given by
\beq
\Phi (\gamma ) =~\sum_{k=0}^{\infty }\frac{(-1)^{k}} {(k+\gamma)^2 }\biggl[\Psi
\left(k+\gamma +1\right)-\Psi (1)\biggr]. \label{9}
\eeq
Using the perturbative expansion of the anomalous dimension one easily determines the leading singularity structure
\beqa
 \gamma&=&\left(2+0\,\omega+\Op(\omega^2)\right)
\left(\frac{-4\,g^2}{\omega}\right) -\left(0+0\,\omega
+\Op(\omega^2)\right)\,\left(\frac{-4\,g^2}{\omega}\right)^2
\nonumber
\\
&&
+\left(0+\,\zt\,\omega +\Op(\omega^2) \right)\,\left(\frac{-4\,g^2}{\omega}\right)^3
-\left(4\,\zt+\frac{5}{4}\,\zfr\,\omega +\Op(\omega^2)\right)\,\left(\frac{-4\,g^2}{\omega}\right)^4 \nonumber \\
&&
-\left(0+\bigg(2\,\ztw\,\zt+16\,\zf\bigg)\,\omega+\Op(\omega^2)\right)\left(\frac{-4g^2}{\omega}\right)^5
\pm \ldots .
\label{BFKLPredictions}
\eeqa

\subsection{Generalized double-logarithmic equation} \label{sec:GenerDL}

Another class of constraints on the anomalous dimension of twist-2 operators is provided by the double-logarithmic asymptotic of the scattering amplitudes.
The double-logarithmic asymptotic of the scattering amplitudes was investigated in QED and QCD in refs.~\cite{Gorshkov:1966ht,Gorshkov:1966hu,Gorshkov:1966qd} and~\cite{Kirschner:1982qf,Kirschner:1982xw,Kirschner:1983di}
(see also {\texttt{arXiv}} version of ref.~\cite{Kotikov:2002ab}). It corresponds to summing the leading terms $(\alpha \ln^2s)^k$ in all orders of perturbation theory.
In combination with a Mellin transformation, the double-logarithmic asymptotic allow to predict the singular part of anomalous dimensions near the point $\M=-2+\omega$.
For our purpose and in our notation the double-logarithmic equation has the following form
\begin{equation}\label{DL}
\gamma\,(2\,\omega+\gamma)=-16\,g^2\,.
\end{equation}
The solution of this equation gives a prediction for the highest pole $(g^{2k}/\omega^{2k-1})$ in all orders of perturbative theory:
\begin{eqnarray}\label{dlevenp}
\gamma&=&-\omega+\omega\, \sqrt{1-\frac{16 g^2}{\omega^2}}
=
2\,\frac{(-4\, g^2)}{\omega}
-2\,\frac{(-4\, g^2)^2}{\omega^3}
+4\,\frac{(-4\, g^2)^3}{\omega^5}
-10\,\frac{(-4\, g^2)^4}{\omega^7}\nonumber \\
&&\hspace*{40mm}
+28\,\frac{(-4\, g^2)^5}{\omega^9}
-84\,\frac{(-4\, g^2)^6}{\omega^{11}}
+\ldots\, .\label{DLSolve}
\end{eqnarray}

The investigation of the analytical properties of the anomalous dimension of twist-2 operators in $\cN=4$ SYM theory led to the suggestion about a simple generalization of the double-logarithmic equation~\cite{Velizhanin:2011pb}\footnote{For the first time, a such generalization was suggested by Lev N. Lipatov and Andrei Onishchenko at 2004,
but was not published. Then, it was improved by Lev N. Lipatov in ref.~\cite{Kotikov:2007cy}.}. The main idea was that in eq.~(\ref{DL}) the corrections to the leading order equation will modify only the right-hand side and that such modification admit besides an expansion over the coupling constant $g^2$ only the appearance of a regular terms over $\omega$ (and, possible, $\gamma$). Substitute the results for the analytical continuation of the anomalous dimension of twist-2 operators near $\M=-2+\omega$ into eq.~(\ref{DL}) we indeed find the following form of the generalized double-logarithmic equation~\cite{Velizhanin:2011pb}
\beqa\label{DLgenerC}
&&\ \gamma\,(2\,\omega+\gamma)=\sum_{k=1}\sum_{m=0}{\mathfrak C}_m^k\,\omega^m\,g^{2k}
\eeqa
where coefficients ${\mathfrak C}_m^k$ can be read directly from the following expression:
\beqa\label{DLgener}
&&16g^2 \Big[-1+\omega+(1+\ztw) \omega ^2+(1-\zt) \omega ^3\nonumber\\&&\qquad\qquad \qquad \qquad \qquad
+(1+\zfr) \omega ^4+(1-\zf) \omega ^5+
(1+\zsx) \omega^6+(1-\zs) \omega^7\Big]\nonumber\\[2mm]&&\quad
+\,g^4 \bigg[
-64 \ztw
+\omega  (128 \ztw+96 \zt)
+\,\omega ^2 (192 \ztw-160 \zt-8 \zfr)\nonumber\\&&\qquad\
+\,\omega ^3 (-256 \ztw \zt+256 \ztw-224 \zt+152 \zfr+360 \zf)\nonumber\\&&\qquad\
+\,\omega ^4 \left(320 \ztw+144 \zt^2-288 \zt+216 \zfr-144 \zf+\frac{58 \zsx}{3}\right)\nonumber\\&&\qquad\
+\,\omega ^5 \big(-384 \ztw \zf+384 \ztw-280 \zt \zfr-352 \zt+280 \zfr-208 \zf+138 \zsx+707 \zs\big)
\bigg]\nonumber\\[2mm]&&\quad
+\,g^6 \bigg[
128 \zt+256 \zfr
+\omega  (1152 \ztw \zt+512 \zt+672 \zfr-960 \zf)\nonumber\\&&\qquad\
+\,\omega ^2 \left(-2688 \ztw \zt-1056 \zt^2+384 \zt+256 \zfr+1504 \zf-\frac{5000 \zsx}{3}\right)\nonumber\\&&\qquad\
+\,\omega ^3 \bigg(-3968 \ztw \zt+6880 \ztw \zf+1760 \zt^2-1072 \zt \zfr-256 \zt+1248 \zfr+4224 \zf\nonumber\\&&\qquad\qquad\quad\
                   +\,\frac{11696 \zsx}{3}-6412 \zs\bigg)
\bigg]\nonumber\\[2mm]&&\quad
+\,g^8 \bigg[
2560 \ztw \zt+384 \zt^2-128 \zf+\frac{1888 \zsx}{3}\nonumber\\&&\qquad\
+\,16\,\omega  \left(288 \ztw \zt-1296 \ztw \zf-184 \zt^2+452\zt \zfr-680 \zf-\frac{4414 \zsx}{3}+553 \zs\right)
\bigg].\qquad\quad
\eeqa
%where coefficients are listed in eq.~(A.2) of Appendix of ref.~\cite{Velizhanin:2011pb}.
This result has a very remarkable consequence. If we solve equation (\ref{DLgenerC}), we have found
\beq\label{DLgenerSolve}
\gamma_{\mathrm{DL}}(\omega)=-\omega+\sqrt{\smash[b]{\omega^2+\sum_{k=1}\sum_{m=0}{\mathfrak C}_m^k\,\omega^m\,g^{2k}}}\,.
\eeq\\[-3mm]
Perturbatively expanding this solution we can predict in all orders of perturbative theory all poles up to $(g^2/\omega^2)^k\omega^{2\ell}$, if we know the $\ell$-loop anomalous dimension (or first $\ell$ orders in right hand side of eq.~(\ref{DLgenerC})).

\subsection{Large $\M$ limit} \label{sec:LargeM}

Third class of constraints for the anomalous dimensions of composite operators in $\cN=4$ SYM theory comes from the consideration of large $\M$ limit. This limit is controlled by ABA and the corresponding predictions can be obtained at any order of perturbative theory. The main statement, interesting for us, that a wrapping corrections can not modify a leading $(\ln \M)$ behavior of large $\M$ limit.

\section{Five loops from constraints} \label{sec:L5}

In this section we will describe in details the methods of the reconstruction of the anomalous dimension of twist-2 operators from the above constraints and a special numerical algorithms. Really, using the generalized Gribov-Lipatov reciprocity we will reconstruct the reciprocity-respecting function $\reciP^{\Wrap}$, rather $\gamma^{\Wrap}$ itself, because in this case the basis will consist of the binomial harmonic sums~(\ref{BinSums}), the number of which is considerable less then the number of the usual harmonic sums~(\ref{vhs}). At five loops $\gamma_{10}^{\Wrap}$ and $\reciP_{10}^{\Wrap}$ are related as
\beq \label{gamma10P}
\gamma_{10}^{\Wrap}=
\frac{1}{2} \cP^{\Wrap}_{8}\dot\gamma_2
+\frac{1}{2}\dot \cP^{\Wrap}_{8} \gamma_2
+\cP_{10}^{\Wrap}\,,
\eeq
where dots over $\gamma_i$ and $\cP_i$ indicate derivatives of the harmonic sums with respect to their indices (see \cite{Dokshitzer:2005bf,Dokshitzer:2006nm}) and
\beqa
&&\reciP_{8}^{\Wrap}=\reciP_{2}^{2}\cT_{8}\,,\\
&&\cT_{8}=\Big(-5\,\zf +2\,\HBS_{2}\,\zt+\left(\HBS_{2,1,2}-\HBS_{3,1,1}\right)\Big)\,,\\
&&\gamma_{2}=\cP_{2}=4\,\HBS_1\,.
\eeqa

To find $\cP_{10}^{\Wrap}$ first of all we should write the full basis from the binomial harmonic sums, which will form the suitable ansatz. At five loops we can write the following general expression for the reciprocity-respecting function $\reciP_{10}^{\Wrap}$:
\beqa
\cP_{10}^{\Wrap}&=&
\HBS_1^2\, \cT_{10}
+c_1\,\HBS_1 \big(\HBS_1\HBS_2-\HBS_{2,1}-\HBS_3\big) \cT_{8}
+c_2\,\HBS_1^4\, \cT_{8}\,.
\label{P10Ansatz}
\eeqa
The appearance of the structure with coefficients $c_1$ and $c_2$ are expected from a suggestion, that the more general integrable system applicable for the computations of the wrapping corrections
represents as the set of (at least) two spin-chains, for which ABA are known (note, that $\HBS_1\HBS_2-\HBS_{2,1}-\HBS_3=\cP_{4}/8$).
During the computations of the wrapping corrections at five loops in ref.~\cite{Lukowski:2009ce} we assumed in advance the appearance of such terms, what allowed to find a final result much faster. Using the maximal transcendentality principle one can find, that at five loops the basis for $\cT_{10}$ will contain $1+1+2+2^3+2^6=76$ binomial harmonic sums (multiplied by $\HBS_1^2$):
\beqa
&\big\{&
{\zs},
\zt^2\, {\HBS}_1,
{\zf}\, {\HBS}_2,
{\underline {{\zf}\, {\HBS}_1^2}},
{\zt}\, {\HBS}_4,
{\zt}\, {\HBS}_{3,1},
{\zt}\, {\HBS}_{2,2},
{\zt}\, {\HBS}_{2,1,1},
{\zt}\, {\HBS}_1\, {\HBS}_3,
{\zt}\, {\HBS}_1\, {\HBS}_{2,1},
{\underline {{\zt}\, {\HBS}_1^2\, {\HBS}_2}},
{\underline {{\zt}\, {\HBS}_1^4}},\nonumber\\&&
{\HBS}_7,
{\HBS}_{6,1},
{\HBS}_{5,2},
{\HBS}_{4,3},
{\HBS}_{3,4},
{\HBS}_{2,5},
{\HBS}_{5,1,1},
{\HBS}_{4,2,1},
{\HBS}_{4,1,2},
{\HBS}_{3,3,1},
{\HBS}_{3,2,2},
{\HBS}_{3,1,3},
{\HBS}_{2,4,1},
{\HBS}_{2,3,2},
{\HBS}_{2,2,3},
{\HBS}_{2,1,4},
\nonumber\\&&
{\HBS}_{4,1,1,1},
{\HBS}_{3,2,1,1},
{\HBS}_{3,1,2,1},
{\HBS}_{3,1,1,2},
{\HBS}_{2,3,1,1},
{\HBS}_{2,2,2,1},
{\HBS}_{2,2,1,2},
{\HBS}_{2,1,3,1},
{\HBS}_{2,1,2,2},
{\HBS}_{2,1,1,3},
\nonumber\\&&
{\HBS}_{3,1,1,1,1},
{\HBS}_{2,2,1,1,1},
{\HBS}_{2,1,2,1,1},
{\HBS}_{2,1,1,2,1},
{\HBS}_{2,1,1,1,2},
{\HBS}_{2,1,1,1,1,1},\nonumber\\&&
{\HBS}_1\, {\HBS}_6,
{\HBS}_1\, {\HBS}_{5,1},
{\HBS}_1\, {\HBS}_{4,2},
{\HBS}_1\, {\HBS}_{3,3},
{\HBS}_1\, {\HBS}_{2,4},
{\HBS}_1\, {\HBS}_{4,1,1},
{\HBS}_1\, {\HBS}_{3,2,1},
{\HBS}_1\, {\HBS}_{3,1,2},
{\HBS}_1\, {\HBS}_{2,3,1},
{\HBS}_1\, {\HBS}_{2,2,2},
{\HBS}_1\, {\HBS}_{2,1,3},
\nonumber\\&&
{\HBS}_1\, {\HBS}_{3,1,1,1},
{\HBS}_1\, {\HBS}_{2,2,1,1},
{\HBS}_1\, {\HBS}_{2,1,2,1},
{\HBS}_1\, {\HBS}_{2,1,1,2},
{\HBS}_1\, {\HBS}_{2,1,1,1,1},
\nonumber\\&&
{\underline {{\HBS}_1^2\, {\HBS}_5}},
{\underline {{\HBS}_1^2\, {\HBS}_{4,1}}},
{\underline {{\HBS}_1^2\, {\HBS}_{3,2}}},
{\underline {{\HBS}_1^2\, {\HBS}_{2,3}}},
{\underline {{\HBS}_1^2\, {\HBS}_{3,1,1}}},
{\underline {{\HBS}_1^2\, {\HBS}_{2,2,1}}},
{\underline {{\HBS}_1^2\, {\HBS}_{2,1,2}}},
{\underline {{\HBS}_1^2\, {\HBS}_{2,1,1,1}}},
\nonumber\\&&
{\underline {{\HBS}_1^3\, {\HBS}_4}},
{\underline {{\HBS}_1^3\, {\HBS}_{3,1}}},
{\underline {{\HBS}_1^3\, {\HBS}_{2,2}}},
{\underline {{\HBS}_1^3\, {\HBS}_{2,1,1}}},
{\underline {{\HBS}_1^4\, {\HBS}_3}},
{\underline {{\HBS}_1^4\, {\HBS}_{2,1}}},
{\underline {{\HBS}_1^5\, {\HBS}_{2}}},
{\underline {{\HBS}_1^7}}
\big\}\,,\label{BasisL5}
\eeqa
where we extracted explicitly $\HBS_1$ from the binomial harmonic sums starting with unity $\HBS_{1,a,b,\ldots}\to\HBS_{1}\,\HBS_{a,b,\ldots}$. Really we will not use for our computations the underlining sums restricted only with the terms, which are not more then the first power of $\HBS_1$.
% ADD
The exclusion of the underlining sums is based on our experience: for the wrapping corrections in the $n+4$-th order we expect the $n$-th power of $\HBS_1$.
However, we may use the basis with all sums in (\ref{BasisL5}), but we will have more constraints, especially for the large  $\M$ limit.

At the next step we take an ansatz for $\cP_{10}^{\Wrap}$ from eqs.~(\ref{P10Ansatz})-(\ref{BasisL5}) multiplied by $c_i$ and compute for it:
\begin{enumerate}
  \item[1)] analytical continuation at $\M=-1+\omega$;
  \item[2)] analytical continuation at $\M=-2+\omega$;
  \item[3)] large $\M$ limit.
\end{enumerate}
All above can be computed with the help of
{\texttt{HARMPOL}}~\cite{Remiddi:1999ew} and {\texttt{SUMMER}}~\cite{Vermaseren:1998uu} packages  for {\texttt{FORM}}~\cite{Vermaseren:2000nd}.

After that we form a set of equations, taking an expressions for the ansatz as a left-hand side and the predictions from the Chapter~\ref{sec:weak} as a right-hand side.
Thus, we obtain the system of the linear equations on the coefficients under the following quantities\footnote{All expressions are available from author upon request.
%Their are available also from
%\href{http://thd.pnpi.spb.ru/~velizh/}{\texttt{http://thd.pnpi.spb.ru/\textasciitilde velizh/}}
}:
\begin{enumerate}
  \item[1)] from BFKL equation~(\ref{BFKLPredictions}) at $\M=-1+\omega$:
\beq
\Big\{
\frac{1}{\omega^9},
\frac{\ztw}{\omega^7},
\frac{\ztw}{\omega^6},
\frac{\zfr}{\omega^5},
\frac{\zf}{\omega^4},
\frac{\ztw\zt}{\omega^4}
\Big\}
\eeq
  \item[2)] from generalized double-logarithmic equation~(\ref{DLgener}) at $\M=-2+\omega$:
\beqa
&&\hspace*{-8mm}\Big\{
\frac{1}{\omega^9},
\frac{1}{\omega^8},
\frac{1}{\omega^7},
\frac{\ztw}{\omega^7},
\frac{1}{\omega^6},
\frac{\ztw}{\omega^6},
\frac{\zt}{\omega^6},
\frac{1}{\omega^5},
\frac{\ztw}{\omega^5},
\frac{\zt}{\omega^5},
\frac{\zfr}{\omega^5},
\frac{1}{\omega^4},
\frac{\ztw}{\omega^4},
\frac{\zt}{\omega^4},
\frac{\zfr}{\omega^4},
\frac{\zf}{\omega^4},
\frac{\ztw\zt}{\omega^4},
\frac{1}{\omega^3},
\frac{\ztw}{\omega^3},
\frac{\zt}{\omega^3},\nonumber\\&&\quad
\frac{\zfr}{\omega^3},
\frac{\zf}{\omega^3},
\frac{\ztw\zt}{\omega^3},
\frac{\zsx}{\omega^3},
\frac{\zt^2}{\omega^3},
\frac{1}{\omega^2},
\frac{\ztw}{\omega^2},
\frac{\zt}{\omega^2},
\frac{\zfr}{\omega^2},
\frac{\zf}{\omega^2},
\frac{\ztw\zt}{\omega^2},
\frac{\zsx}{\omega^2},
\frac{\zt^2}{\omega^2},
\frac{\zs}{\omega^2},
\frac{\ztw\zf}{\omega^2},
\frac{\zt\zfr}{\omega^2}\Big\}
\eeqa
  \item[3)] from large $\M$ limit:
\beq
\Sinf^2\Big\{\zs,\ztw\zf,\zfr\zt,\Sinf\zsx,\Sinf\zt^2,\Sinf^2\zf,\Sinf^2\ztw\zt\Big\},\qquad \Sinf=S_1(\infty)
\eeq
\end{enumerate}
So, we have the system from $49$ equations with $59$ unknowns. However, some of them are linear depended. Remove such equations, we are left with $40$ equations on $59$ unknowns.

As was pointed out in Introduction the obtained equations have a rather special form~- they give rise the system of linear Diophantine equations, because we are believe, that the coefficients in ansatz, which we are looking for, are an integer numbers. To solve this problem we will address to two methods from the numbers theory: LLL-algorithm and Linear Programming.

\subsection{LLL-algorithm} \label{sec:LLL}

The Lenstra-Lenstra-Lov\'asz (LLL) %lattice basis reduction
algorithm is a polynomial time lattice reduction algorithm invented by Arjen Lenstra, Hendrik Lenstra and L\'aszl\'o Lov\'asz in 1982~\cite{Lenstra:1982}.
%Given a basis \mathbf{B}=\{ \mathbf{b}_1,\mathbf{b}_2, \dots, \mathbf{b}_d \} with n-dimensional integer coordinates, for a lattice L in Rn with  \ d \leq n , the LLL algorithm outputs an LLL-reduced (short, nearly orthogonal) lattice basis in time O(d^5n\log^3 B)\, where B is the largest length of b_i under the Euclidean norm.
The original applications were to give polynomial time algorithms for factorizing polynomials with rational coefficients into irreducible polynomials, for finding simultaneous rational approximations to real numbers, and for solving the integer linear programming problem in fixed dimensions.

Let's give step-by-step description how to apply the LLL-algorithm to the solution of our problem on a toy example.
This example is related with the rational part of the wrapping corrections for the four-loop anomalous dimension of twist-2 operators, which was calculated in ref.~\cite{Bajnok:2008qj}. That is, we will try to obtain the expression
\beq
\Delta^{\mathrm {\Wrap,\,rational}}_8(\M)={\HS}_{-2,1}-\frac{{\HS}_{-3}}{2}-{\HS}_{-2} {\HS}_1-{\HS}_1 {\HS}_2-\frac{{\HS}_3}{2}
\label{L4Wraprat}
\eeq
from the values at $\M=1$ and at $\M=2$ for the following basis with the binomial harmonic sums
\beq
\left\{{\HBS}_{3,2},{\HBS}_{2,3},{\HBS}_{2,2,1},{\HBS}_{2,1,2},{\HBS}_{3,1,1}\right\}\,.
\eeq
In this basis eq.~(\ref{L4Wraprat}) is looks like:
\beq
\Delta^{\mathrm {\Wrap,\,rational}}_8(\M)=-{\HBS}_{2,1,2}+{\HBS}_{3,1,1}\,.
\eeq
We will use {\texttt{LatticeReduce}} function of {\texttt{MATHEMATICA}}, which realize the LLL-algorithm\footnote{See {\texttt{Application}} on
\href{http://reference.wolfram.com/mathematica/ref/LatticeReduce.html}
{\texttt{{http://reference.wolfram.com/mathematica/ref/LatticeReduce.html}}}}.
% ADD
Strictly speaking, the LLL-algorithm try to reduce an original lattice to the lattice in which all rows (each row is a vector on the lattice) will have the shortest Euclidian norm.
\begin{itemize}
\item
We start from the system of equations:
\beqa
 x_1 + x_2 + x_3 + x_4  + x_5&=& 0\nonumber\\
\frac{15}{32}\,x_1 + \frac{27}{32}\,x_2 + \frac{33}{32}\,x_3 + \frac{39}{32}\,x_4 + \frac{21}{32}\,x_5&=&\frac{9}{16}
\label{OrigSysEqs}
\eeqa
\item
take matrix from the coefficients of above system of equations and right-hand side term as last columns
\item
divide each row of above matrix to the greatest common divisor ({\texttt{GCD}} function):
\beq
{\mathsf{SE}}=
\left(
\begin{array}{cccccc}
 1 & 1 & 1 & 1 & 1 & 0 \\
 5 & 9 & 11 & 13 & 7 & -6
\end{array}
\right)
\label{SE}
\eeq
\item
multiply ${\mathsf{SE}}$ to some huge integer number, for example $8^8$
\item
create an unity matrix $\mathbb I$ with a rank equal to the length of rows in ${\mathsf{SE}}$
\item
append transpose ${\mathsf{SE}}$
to the right side of the  unity matrix $\mathbb I$:
\beq
%{\mathbb{I}\mathsf{SE}}=
\left(
\begin{array}{cccccccr}
 1\ &\ 0\ &\ 0\ &\ 0\ &\ 0\ &\ 0\ &\ 8^8\ &\ 5  \times 8^8 \\
 0\ &\ 1\ &\ 0\ &\ 0\ &\ 0\ &\ 0\ &\ 8^8\ &\ 9  \times 8^8 \\
 0\ &\ 0\ &\ 1\ &\ 0\ &\ 0\ &\ 0\ &\ 8^8\ &\ 11 \times 8^8 \\
 0\ &\ 0\ &\ 0\ &\ 1\ &\ 0\ &\ 0\ &\ 8^8\ &\ 13 \times 8^8 \\
 0\ &\ 0\ &\ 0\ &\ 0\ &\ 1\ &\ 0\ &\ 8^8\ &\ 7  \times 8^8 \\
 0\ &\ 0\ &\ 0\ &\ 0\ &\ 0\ &\ 1\ &\ 0  \ &\ -6 \times 8^8
\end{array}
\right)
\label{ISE}
\eeq
\item
apply {\texttt {LatticeReduce}} to this matrix
\end{itemize}
As result we obtain the following matrix:
\beq
{\mathsf{RSE}}=
\left(
\begin{array}{cccccccr}
 -1\ &\ 0\ &\ 1 \ &\ -1\ &\ 1 \ &\ 0 \ &\ 0 \   &\ 0\ \; \\
 0 \ &\ 0\ &\ 0 \ &\ 1 \ &\ -1\ &\ 1 \ &\ 0 \   &\ 0\ \; \\
 -1\ &\ 1\ &\ -1\ &\ 0 \ &\ 1 \ &\ 0 \ &\ 0 \   &\ 0\ \; \\
 0 \ &\ 1\ &\ 1 \ &\ -1\ &\ -1\ &\ 0 \ &\ 0 \   &\ 0\ \; \\
 0 \ &\ 0\ &\ 0 \ &\ 0 \ &\ -1\ &\ -1\ &\ -8^8\ &\ -8^8 \\
 -1\ &\ 0\ &\ 0 \ &\ 0 \ &\ 0 \ &\ -1\ &\ -8^8\ &\ 8^8
\end{array}
\right)
\label{RSE}
\eeq
% ADD
Multiplication of eq.~(\ref{SE}) to some huge integer number makes the huge Euclidian norms for all vectors in eq.~(\ref{ISE}), what allows the LLL-algorithm more effectively find the vectors with the shortest  Euclidian norms in eq.~(\ref{RSE}).
We are interesting only with the rows, which are non zero at position 6 (number of variables plus one - such column corresponds to the right hand sides or free terms in eq.~(\ref{OrigSysEqs})) and with the rest numbers at right equal to zeros.
% ADD
Such vector should exist always, because it solves the original system of equations~(\ref{OrigSysEqs}), while other vectors in ${\mathsf{RSE}}$~(\ref{RSE}) without large numbers solve the homogeneous system of the original system of equations~(\ref{OrigSysEqs}).
Only row 2 satisfies above criteria. Indeed, this is the solution, which we were looking for.
% ADD
Because the rows 1, 3 and 4 are the solutions of the homogeneous system of the original system of equations~(\ref{OrigSysEqs}) these rows can be added to the row 2 and an obtained solution is also the solution of the original system of equations~(\ref{OrigSysEqs}). In matrix (\ref{RSE}) the rows arranged according their Euclidian norms, so a criterion for the correctness of the solution is a position of the corresponding row in the matrix, which will be obtained after the application of LLL-algorithm: the higher the row, the more plausible solution.

Note one important property of the LLL-algorithm (and all other similar algorithms): a desired numbers should be small, at least most of them, because LLL-algorithm is looking for the vectors with minimal Euclidean norm.
Therefore, for the reconstruction of the five-loop planar anomalous dimension of twist-2 operators we should eliminate some possible large numbers from the result, which we are looking for. From the four-loop anomalous dimension of twist-2 operators~\cite{Bajnok:2008qj} and the five-loop anomalous dimension of twist-3 operators~\cite{Beccaria:2009eq} we can find such large numbers as the coefficients in the front of terms in the basis with zeta-numbers $\zeta_i$.
To avoid possible difficulties we can eliminate the variables from the system of equations, which should be a large numbers as we believe.
However we should keep as much equations as possible because each equation preserves an information that its solution should be an integer numbers, i.e. if we exclude some of variables we can obtain after solution that these variables will not integers.
First of all we should eliminate the variable related with $\HBS_1^2\zs$ term from the basis~(\ref{BasisL5}) as the most dangerous. Moreover, because we do not know overall normalization of the result, we can try a different variants for the normalization, for example in the form $2^{-i}$, which will a prefactor of the right hand side of our equations (or we can rescale the binomial harmonic sums~(\ref{BinSums})).
For $i=9$ we obtain the solution with rather small sum of absolute values and with a lot of zeros
\beqa
&\{&\!4, 1, %105,
-6, -40, -4, 8, -8, -4, 0, 12, 0, 0, 0, 0, 0, 0, 2, 0, -2, 4, 4, -10, 0, 0, 4, -2, 0, 4, 4, -4, 4, \nonumber\\
&& -4, -4, 4, -4, 0, 2, 0, 0, 0, -2, 0, 0, 0, 0, 0, 0, 0, 0, -2, 2, 0, 0, 0, 0, 0, 0, 0, 16, 0,\ldots
\}\,,\qquad\quad
\eeqa
where first two numbers corresponds $c_1$ and $c_2$ and the rest numbers are listed according to the basis~(\ref{BasisL5}) with missing $\HBS_1^2\zs$ term, which can be obtained from one of the initial equations (it equals $105$).
Indeed, this solution (multiplied by common factor $2^{9-4}=32$) gives the result, which is the same as was obtained earlier in ref.~\cite{Lukowski:2009ce} with the calculation of the finite size effects
\beqa \label{gamma10split}
\cP_{10}^{\Wrap}&=&2\,\cP_2^2 \cT_{10}
+2\,\cP_2 \Big(2\,\cP_4 +\frac{1}{16} \cP^3_2\Big)
\Big(-5\,\zf +2\,\HBS_{2}\,\zt+\left(\HBS_{2,1,2}-\HBS_{3,1,1}\right)\Big)\,,\\[2mm] \label{Ttilde}
\cT_{10}&=&
105\,\zs
-6\,\HBS_1\,{\zt}^2
-40\,\HBS_2\,\zf
+4\big( 3\,\HBS_1 \HBS_{2,1} -2\,\HBS_{2,2}+2\,\HBS_{3,1}-\HBS_{2,1,1}-\HBS_4\big)\,\zt\nonumber\\
&+&2 \, \Big(
\HBS_1 \left(\HBS_{2,3,1}-\HBS_{3,1,2}\right)
-\HBS_{2,1,4}
+2\, \HBS_{2,2,3}
-5\, \HBS_{3,1,3}
+2\, \HBS_{3,2,2}\nonumber\\
&&\quad\ +2\, \HBS_{3,3,1}
-\HBS_{4,1,2}
+\HBS_{5,1,1}
-2\, \HBS_{2,1,2,2}
+2\, \HBS_{2,1,3,1}
-2\, \HBS_{2,2,1,2}
-2\, \HBS_{2,2,2,1}\nonumber\\
&&\quad\ +2\, \HBS_{2,3,1,1}
-2\, \HBS_{3,1,1,2}
+2\, \HBS_{3,1,2,1}
+2\, \HBS_{3,2,1,1}
-\HBS_{2,1,1,1,2}
+\HBS_{3,1,1,1,1}\Big)\,.
\eeqa

In this way we get the result for the wrapping effect part of the five-loop planar anomalous dimension of twist-2 operators without any calculations at all, using only the properties of harmonic sums and the available information, obtained from the already known results from the low orders of perturbative theory.

\subsection{Linear Programming} \label{sec:MIP}

We have found also other method, which can be applied for the solution of the system of linear Diophantine equations. As we pointed out the LLL-algorithm seeking the vectors with the minimal Euclidean norms. However, it seems, that the more correct criteria is a minimization of the sum of absolute values of coefficients instead of the sums of their squares. In this case our problem is reduced to Integer Linear Problem, which is particular case of Linear Programming or Linear Optimization.
Formally, linear programming is a technique for the optimization of a linear objective function, subject to linear equality and linear inequality constraints.
We have a lot of constraints (equations) and want to find the best solution, which satisfied some global condition, in our case - the minimum of the sum of the absolute values. {\texttt {MATHEMATICA}} has a functions, which are related with an optimization. Most simple is {\texttt {Minimize}}. Application of {\texttt {Minimize}} is rather straightforward and we will not describe it here.

In general, {\texttt {Minimize}} function tries to find the global minimum in the multidimensional space (equal to the number of variables) under the required conditions and then move to the nearest integer solution. In our case we will have 58-dimensional space and such task can not be solved with {\texttt{MATHEMATICA}} in a reasonable time. However we have found some interesting future of the solution for the five-loop planar anomalous dimension of twist-2 operators. In principal, we can use some additional information about anomalous dimension. Namely, we used the result for the one impurity state for the twist-2 operators in the $\beta$-deformed $\cN=4$ SYM theory, which are known even up to six-loop order~\cite{Bajnok:2008qj,Bajnok:2009vm,Bajnok:2010ud}. This result corresponds to the value of the anomalous dimension of the twist-2 operators in $\cN=4$ SYM theory at $\M=1$. Excluding the variables related with $\HBS_1^2\zs$ and $\HBS_1^2\zf\HBS_2$ terms from the basis~(\ref{BasisL5}) and solving the most simple equations with these variables we obtain the system in which, as we believe, all unknowns should be less then $2^4=16$ (we also rescaled all binomial harmonic sums in basis~(\ref{BasisL5}) by factor $2^5=32$). Run {\texttt {NMinimize}} under these conditions we have found, that the obtained {\it numerical} solution is much close to the solution, which we should obtain. In other words the global minimum of the multidimensional problem lies near the exact solution!

Because the Linear Optimization is very actual at the present time there are a lot of specials programs, which realize this procedure in a rapid way with the usage of special algorithm and which can be run under modern computer clusters with parallelization.
Usually Linear Optimization works with a positive integer quantities and an available programs are restricted only to the positive integer numbers. But because we do not know an exact signs of the coefficients in ansatz we should symmetrize the corresponding variables, doubling the number of unknowns. It would be very interesting to find a such symmetry, which can fix these signs.

\section{Conclusion and discussion} \label{sec:discussion}

In this paper we reconstruct the full planar five-loop anomalous dimension of twist-2 operators in $\cN=4$ SYM theory using the known constraints from the BFKL equation, the generalized double-logarithmic equations and from the large spin limit with the help of numerical methods and without any special computations.

In principal described methods can be used for the solution of a similar problems, where desired results are looking in a known basis and some information about the properties of these results are available.
In spite of simplicity the LLL-algorithm gives a sought-for result in a more simple way. As an example of its application we have reconstructed the next-to-leading order wrapping correction to the anomalous dimension of the twist-2 operators in the $\beta$-deformed $\cN=4$ SYM theory from the results of ref.~\cite{deLeeuw:2010ed}, which can be find in Appendix~\ref{App}. Note, however, that LLL-algorithm can not find a desired solution if the numbers in matrix ${\mathsf{SE}}$ in eq.~(\ref{SE}) are rather small or the rank of the system of equations is much less, then the numbers of variables.
Linear Programming is more exact method, but more time-consuming. We used its for the reconstruction of the non-planar contribution to the four-loop anomalous dimension of the twist-2 operators in $\cN=4$ SYM theory from our results~\cite{Velizhanin:2009gv,Velizhanin:2010ey} and some additional constraints, because the usage of LLL-algorithm is beyond its applicability\footnote{We have found some reasonable solutions, but we are going to check their with the forthcoming direct perturbative calculations.}.
Probably, we should combine these two methods using an output of LLL-algorithm as an input (first approximation or initial solution) of linear programming methods.

It seems, that at six loops an available information is not enough, because the basis from the binomial harmonic sums growth faster, then the number of constraints. There are some possibilities, which can help to solve this problem. First of all we can used some information, which is available for low values of $\M$: six-loop anomalous dimension for one-impurity state in $\beta$-deformed $\cN=4$ SYM theory~\cite{Bajnok:2010ud}, six-loop anomalous dimension for Konishi~\cite{Bajnok:2012bz,Leurent:2012ab,Leurent:2013mr} and other, but such information will be based on the methods, which we do not want to address. Among another possibilities we may try to find an additional symmetry, which can reduce the basis, either extend the number of constraints, for example, from the studying of large $\M$ limit or an expansion near $\M=0$~\cite{Basso:2011rs}.

%%%%%%%%%%%%%%%%%%%%%%%%%%%%%%%%%%%%%%%%%%%%%%%%%%%%%%%%%%%%%%%%%%

\acknowledgments

I would like to thank to Andrei Onishchenko and Lev Nikolaevich Lipatov for the fruitful and stimulated discussions. %This work is supported
This research is supported by a Marie Curie International Incoming Fellowship within the 7th European Community Framework Programme, grant number PIIF-GA-2012-331484 and by RFBR grants 13-02-01246-a, RSGSS-4801.2012.2.

%%%%%%%%%%%%%%%%%%%%%%%%%%%%%%%%%%%%%%%%%%%%%%%%%%%%%%%%%%%%%%%%%%
\newpage
\appendix
%%%%%%%%%%%%%%%%%%%%%%%%%%%%%%%%%%%%%%%%%%%%%%%%%%%%%%%%%%%%%%%%%%

\section{NLO wrapping corrections in the $\beta$-deformed $\cN=4$ SYM theory} \label{App}

In this Appendix we give a general expression for the next-to-leading order (NLO) wrapping correction to the anomalous dimension of twist-2 operators in the $\beta$-deformed $\cN=4$ SYM theory from the results of ref.~\cite{deLeeuw:2010ed} reconstructed with the help of LLL-algorithm. The wrapping corrections in the leading  order for twist-2 operators can be written as (see eq.~(4.5) in ref.~\cite{deLeeuw:2010ed}):
\beq\label{SimpleTwist2}
E^{\Wrap}_{{\mathrm{LO}}}(\M) = 4g^6 \sin^2 (2\pi\beta)\frac{\HS_1(\M)\HS_{-2}(\M)}{\M(\M+1)}
\eeq
The results for the wrapping corrections in the next-to leading order can be find in the table from ref.~\cite{deLeeuw:2010ed} for the first ten even values of $\M$. To reconstruct a general expression we write down the following basis:
\beqa
&\bigg\{&\HS_1\frac{\HS_1\HS_{-2}}{\M},\HS_1\frac{\HS_1\HS_{-2}}{\M+1},\HS_1 \HS_2\HS_{-2},\HS_1 \HS_1 \HS_{-3},
   {\HBS}_5,{\HBS}_{4,1},{\HBS}_{3,2},{\HBS}_{2,3},{\HBS}_{3,1,1},{\HBS}_{2,2,1},{\HBS}_{2,1,2},{\HBS}_{2,1,1,1},\nonumber\\
&&   {\HBS}_1 {\HBS}_4,{\HBS}_1 {\HBS}_{3,1},{\HBS}_1 {\HBS}_{2,2},{\HBS}_1 {\HBS}_{2,1,1},{\HBS}_1^2 {\HBS}_3,
   {\HBS}_1^2 {\HBS}_{2,1}\bigg\}\frac{1}{\M(\M+1)}\,,
\eeqa
where first four terms come from the reciprocity $(\gamma_0\,E^{\Wrap}_{{\mathrm{LO}}})\,'$. We need only four first values from the table of ref.~\cite{deLeeuw:2010ed} to obtain with the help of {\texttt{LatticeReduce}} functions the following vector:
\beq
\{16, 0, 16, 16, 0, -1, -1, 0, -2, 1, -1, 0, -2, 4, -2, 0, 1, 0, 4, 0, 0, 0, 0\}
\eeq
which give the desired result:
\beqa
&&E^{\Wrap}_{{\mathrm{NLO}}}(\M) = g^8 \sin^2 (2\pi\beta)\frac{1}{4\,\M(\M+1)}\bigg[
16\bigg(\HS_1\frac{\HS_1\HS_{-2}}{\M}+\HS_1 \HS_2\HS_{-2}+\HS_1 \HS_1 \HS_{-3}\bigg)\nonumber\\
&&\qquad - {\HBS}_{4,1} - {\HBS}_{3,2}
%,{\HBS}_{2,3},
-2\,{\HBS}_{3,1,1}
+{\HBS}_{2,2,1}
-{\HBS}_{2,1,2}
%,{\HBS}_{2,1,1,1},\nonumber\\
-2\,{\HBS}_1 {\HBS}_4
+4\,{\HBS}_1 {\HBS}_{3,1}
-2\,{\HBS}_1 {\HBS}_{2,2}
%,{\HBS}_1 {\HBS}_{2,1,1}
+{\HBS}_1^2 {\HBS}_3
%   {\HBS}_1^2 {\HBS}_{2,1}
\bigg].\qquad
\eeqa
One can easily check with other values from the table of ref.~\cite{deLeeuw:2010ed} that this is indeed the correct answer.

%%%%%\bibliographystyle{JHEP}
%%%%%
%%%%%\bibliography{1311.6953v2}
%%%%%
%%%%%\end{document}
\newpage

\providecommand{\href}[2]{#2}\begingroup\raggedright

\endgroup

%%%%%%%%%%%%%%%%%%%%%%%%%%%%%%%%%%%%%%%%%%%%%%%%%%%%%%%%%%%%
\end{document}